\documentclass{vgtc}

\ifpdf
  \pdfoutput=1\relax
  \usepackage{graphicx}
  \DeclareGraphicsExtensions{.pdf,.png,.jpg,.jpeg}
\else
  \usepackage{graphicx}
  \DeclareGraphicsExtensions{.eps}
\fi

\graphicspath{{figures/}{pictures/}{images/}{./}}

\usepackage{microtype}
\PassOptionsToPackage{warn}{textcomp}
\usepackage{textcomp}
\usepackage{mathptmx}
\usepackage{times}

\usepackage{cite}
\usepackage{tabu}
\usepackage{booktabs}

\usepackage{tikz}
\usetikzlibrary{arrows,arrows.meta,patterns,calc,shapes,decorations.pathreplacing}
\newcommand{\boundellipse}[3]
{(#1) ellipse (#2 and #3)
}

\definecolor{branchColor1}{HTML}{1F77B4} 
\definecolor{branchColor2}{HTML}{AEC7E8} 
\definecolor{branchColor3}{HTML}{2CA02C} 
\definecolor{branchColor4}{HTML}{98DF8A} 
\definecolor{branchColor7}{HTML}{D62728} 
\definecolor{branchColor5}{HTML}{FFBB78} 
\definecolor{branchColor6}{HTML}{FF7F0E} 
\definecolor{branchColor8}{HTML}{9C00FF} 
\definecolor{branchColor9}{HTML}{D189FF} 

\definecolor{grayC}{HTML}{CCCCCC}
\definecolor{grayB}{HTML}{BBBBBB}
\definecolor{grayA}{HTML}{AAAAAA}
\definecolor{gray9}{HTML}{999999}
\definecolor{gray8}{HTML}{888888}
\definecolor{gray7}{HTML}{777777}
\definecolor{gray6}{HTML}{666666}
\definecolor{gray5}{HTML}{555555}
\definecolor{gray4}{HTML}{444444}

\definecolor{filtered}{HTML}{BBBBBB}
\definecolor{filtered2}{HTML}{CCCCCC}
\definecolor{filtered3}{HTML}{F5F5F5}

\definecolor{lm2}{HTML}{6FB0E7}
\definecolor{lm1}{HTML}{2484D6}
\definecolor{l1}{HTML}{BD0026}
\definecolor{l2}{HTML}{F03B20}
\definecolor{l3}{HTML}{FD8D3C}
\definecolor{l4}{HTML}{FED976}
\definecolor{l5}{HTML}{FFFFB2}

\definecolor{colorBrewerC12_0}{HTML}{a6cee3}
\definecolor{colorBrewerC12_1}{HTML}{1f78b4}
\definecolor{colorBrewerC12_2}{HTML}{b2df8a}
\definecolor{colorBrewerC12_3}{HTML}{33a02c}
\definecolor{colorBrewerC12_4}{HTML}{fb9a99}
\definecolor{colorBrewerC12_5}{HTML}{e31a1c}
\definecolor{colorBrewerC12_55}{HTML}{a40013}
\definecolor{colorBrewerC12_6}{HTML}{fdbf6f}
\definecolor{colorBrewerC12_7}{HTML}{ff7f00}
\definecolor{colorBrewerC12_8}{HTML}{cab2d6}
\definecolor{colorBrewerC12_9}{HTML}{6a3d9a}
\definecolor{colorBrewerC12_10}{HTML}{ffff99}
\definecolor{colorBrewerC12_11}{HTML}{b15928}
\definecolor{colorBrewerC12_12}{HTML}{595959}
\definecolor{colorBrewerC12_13}{HTML}{e80500}

\definecolor{ng3dt1}{HTML}{3182bd}
\definecolor{ng3dt2}{HTML}{6baed6}
\definecolor{ng3dt3}{HTML}{9ecae1}

\definecolor{ng3dn1}{HTML}{cc0000}
\definecolor{ng3dn2}{HTML}{d96666}
\definecolor{ng3dn3}{HTML}{ffabab}

\tikzstyle{contourNumber0} = [fill=white, draw=black, circle, inner sep=0pt]
\tikzstyle{contourNumber1} = [fill=white, draw=black, circle, inner sep=0.5pt]
\tikzstyle{contourNumber2} = [fill=white, draw=black, circle, inner sep=0.04em]

\tikzstyle{veryThinEdge} = [line width=0.01cm]
\tikzstyle{thinEdge} = [line width=0.03cm]
\tikzstyle{mediumEdge} = [line width=0.04cm]
\tikzstyle{thickEdge} = [line width=0.1cm]
\tikzstyle{dashedEdge} = [dashed, gray]

\tikzstyle{thickNode} = [shape=circle,draw=black,fill=black, inner sep=0pt, minimum size=4]
\tikzstyle{filteredNode} = [shape=circle,draw=filtered,fill=filtered, inner sep=0pt, minimum size=4]

\tikzstyle{arrow} = [-{Stealth[scale=1.5]}]
\tikzstyle{arrow2} = [-{Stealth[scale=1.1]}]

\definecolor{rLevel0}{HTML}{a93838}
\definecolor{rLevel1}{HTML}{e3aeae}
\definecolor{bLevel0}{HTML}{3e7fa4}
\definecolor{bLevel1}{HTML}{b1cfe1}
\definecolor{gLevel0}{HTML}{3ea440}
\definecolor{gLevel1}{HTML}{b1e1b2}

\definecolor{colorTheme_00}{HTML}{a71d44}
\definecolor{colorTheme_01}{HTML}{dd6767}

\definecolor{colorTheme_10}{HTML}{777777}
\definecolor{colorTheme_11}{HTML}{bbbbbb}

\definecolor{colorTheme_20}{HTML}{f27d00}
\definecolor{colorTheme_21}{HTML}{f1b473}
\definecolor{colorTheme_22}{HTML}{edc293}
\definecolor{colorTheme_23}{HTML}{f5cca1}

\definecolor{colorTheme_30}{HTML}{3ea440}

\definecolor{colorTheme_40}{HTML}{1a769c}
\definecolor{colorTheme_41}{HTML}{83a6c1}

\definecolor{simplexBG}{HTML}{ececec}

\definecolor{highlight0_r}{HTML}{e3aeae}
\definecolor{highlight1_r}{HTML}{a93838}

\definecolor{highlight0_b}{HTML}{b1cfe1}
\definecolor{highlight1_b}{HTML}{3e7fa4}

\tikzstyle{simplexVertex}=[draw=black, thinEdge, fill=white, shape=circle, minimum size=7pt, inner sep=2pt]
\tikzstyle{highlightedSimplexVertexR}=[draw=colorTheme_00, line width=0.75mm, fill=colorTheme_01, shape=circle, minimum size=1.3em, inner sep=2pt]
\tikzstyle{highlightedSimplexVertexB}=[draw=colorTheme_20, line width=0.75mm, fill=colorTheme_23, shape=circle, minimum size=1.3em, inner sep=2pt]
\tikzstyle{highlightedSimplexVertexG}=[draw=colorTheme_10, line width=0.75mm, fill=colorTheme_11, shape=circle, minimum size=1.3em, inner sep=2pt]

\tikzstyle{simplexEdge}=[draw=black, line width=0.1mm]
\tikzstyle{highlightedSimplexEdgeR}=[draw=colorTheme_00, thickEdge]
\tikzstyle{highlightedSimplexEdgeB}=[draw=colorTheme_20, thickEdge]
\tikzstyle{highlightedSimplexEdgeG}=[draw=colorTheme_10, thickEdge]

\tikzstyle{simplexTriangle}=[fill=simplexBG, draw=black, line width=0.1mm]
\tikzstyle{highlightedSimplexTriangleR}=[fill=rLevel1, draw=black, line width=0.1mm]
\tikzstyle{highlightedSimplexTriangleB}=[fill=bLevel1, draw=black, line width=0.1mm]
\tikzstyle{highlightedSimplexTriangleG}=[fill=gLevel1, draw=black, line width=0.1mm]

\tikzstyle{persistenceCurveMarker} = [circle, draw=rLevel0, line width=2pt, inner sep=0.2em]
\tikzstyle{persistenceCurveArrow} = [draw=black,arrow,densely dashed, line width=0.6pt]

\usepackage{enumitem}

\onlineid{1024}

\vgtccategory{Research}

\vgtcpapertype{System}

\vgtcinsertpkg

\title{Cinema Darkroom:\\A Deferred Rendering Framework for Large-Scale Datasets}

\usepackage{soul}

\newcommand{\edit}[1]{{#1}}

\newcommand{\delete}[1]{}

\author{
Jonas Lukasczyk\thanks{e-mail: jl@jluk.de\vspace*{-3.9em}}\\
     \scriptsize Arizona State University
\and Christoph Garth\\
     \scriptsize TU Kaiserslautern
\and Matthew Larsen\\
     \scriptsize Lawrence Livermore National Laboratory
\and Wito Engelke\\
     \scriptsize Linköping University\vspace*{1em}
\and Ingrid Hotz\\
     \scriptsize Linköping University
\and David Rogers\\
     \scriptsize Los Alamos National Laboratory
\and James Ahrens\\
     \scriptsize Los Alamos National Laboratory
\and Ross Maciejewski\\
     \scriptsize Arizona State University
}

\shortauthortitle{Lukasczyk \MakeLowercase{\textit{et al.}}: Cinema Darkroom}

\abstract{

    \vspace*{-0.3em}This paper presents a framework that fully leverages the advantages of a deferred rendering approach for the interactive visualization of large-scale datasets.
    Geometry buffers (G-Buffers) are generated and stored in situ, and shading is performed post hoc in an interactive image-based rendering front end.
    This decoupled framework has two major advantages.
    First, the G-Buffers only need to be computed and stored once---which corresponds to the most expensive part of the rendering pipeline.
    Second, the stored G-Buffers can later be consumed in an image-based rendering front end that enables users to interactively adjust various visualization parameters---such as the applied color map or the strength of ambient occlusion---where suitable choices are often not known a priori.
    This paper demonstrates the use of Cinema Darkroom on several real-world datasets, highlighting CD's ability to effectively decouple the complexity and size of the dataset from its visualization.

}


\teaser{
    \vspace*{-1em}
    \begin{center}
        \includegraphics[width=0.92\textwidth]{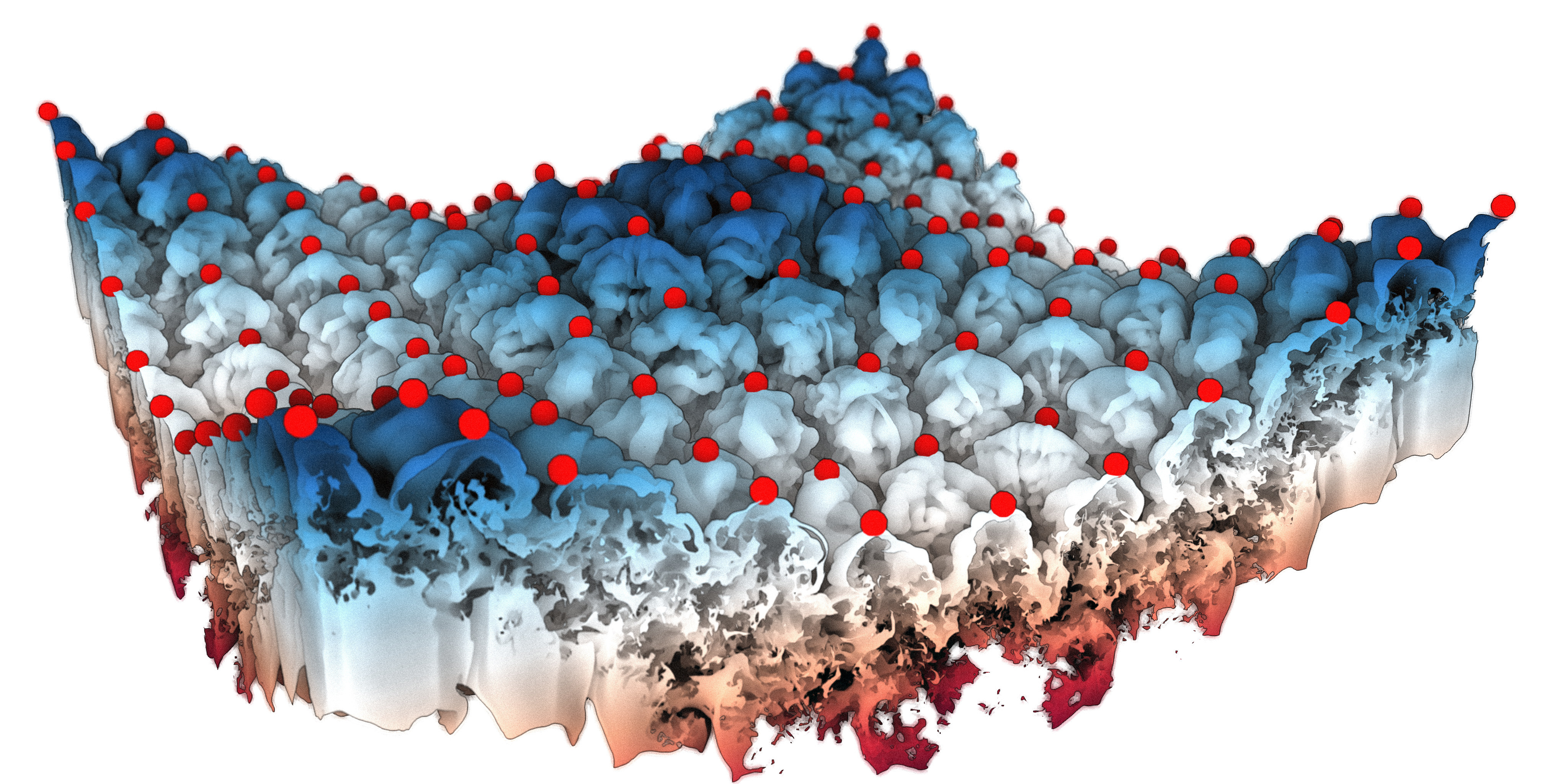}
    \end{center}
    \vspace*{-1.75em}

    \caption{
        Post hoc Cinema Darkroom (CD) rendering of the Richtmyer-Meshkov Instability~(RMI) dataset~\cite{openSciVisDataSets} showing a contour of the entropy field for isovalue $20$ colored by height (colored surface), as well as significant maxima of the height field (red spheres).
        RMI describes the impulsive acceleration of two fluids with different density, leading to the formation of an instability at the fluids' interface~\cite{cohen2002three}.
        The depicted scene consists of roughly $300$ million triangles, but CD utilizes a deferred rendering approach~\cite{deering1988triangle,liang2000deferred} to interactively render the scene based only on a single $4096${}$\times${}$2048$ depth image (for the contour) and a set of $\sim${}$100$ points (for the maxima).
        The depth image, the corresponding camera parameters, and the maxima coordinates have been computed and stored in situ in a Cinema database~\cite{ahrens2015image} via an extension of the Topology ToolKit~\edit{(TTK)}~\cite{ttk2017}.
        CD features a web-based deferred rendering front end (\autoref{fig_frontend}) that consumes these data products for post hoc visualization.
        Since the camera parameters are known, CD can compute the height field in image space and can correctly project spheres at the maxima locations.
        Moreover, CD supports many image-based rendering techniques; including post hoc color mapping~\cite{rogers2018cinemaspec}, depth compositing~\cite{dntg2020}, screen space ambient occlusion~(SSAO)~\cite{ritschel2009approximating}, screen space depth of field~(SSDoF)~\cite{mcintosh2012efficiently}, and fast approximate anti-aliasing~(FXAA)~\cite{lottes2009fast}.
    }
    \label{fig_teaser}
}

\begin{document}

\newcommand{\mycaption}[1]{\vspace*{-1em}\caption{#1}\vspace{-0.25em}}

\renewcommand{\sectionautorefname}{Sec.}
\renewcommand{\subsectionautorefname}{Sec.}
\renewcommand{\subsubsectionautorefname}{Sec.}
\renewcommand{\figureautorefname}{Fig.}
\renewcommand{\equationautorefname}{Eq.}
\renewcommand{\tableautorefname}{Tab.}

\firstsection{Introduction}
\label{sec_introduction}
\maketitle

\begin{figure*}
    \includegraphics[width=\textwidth]{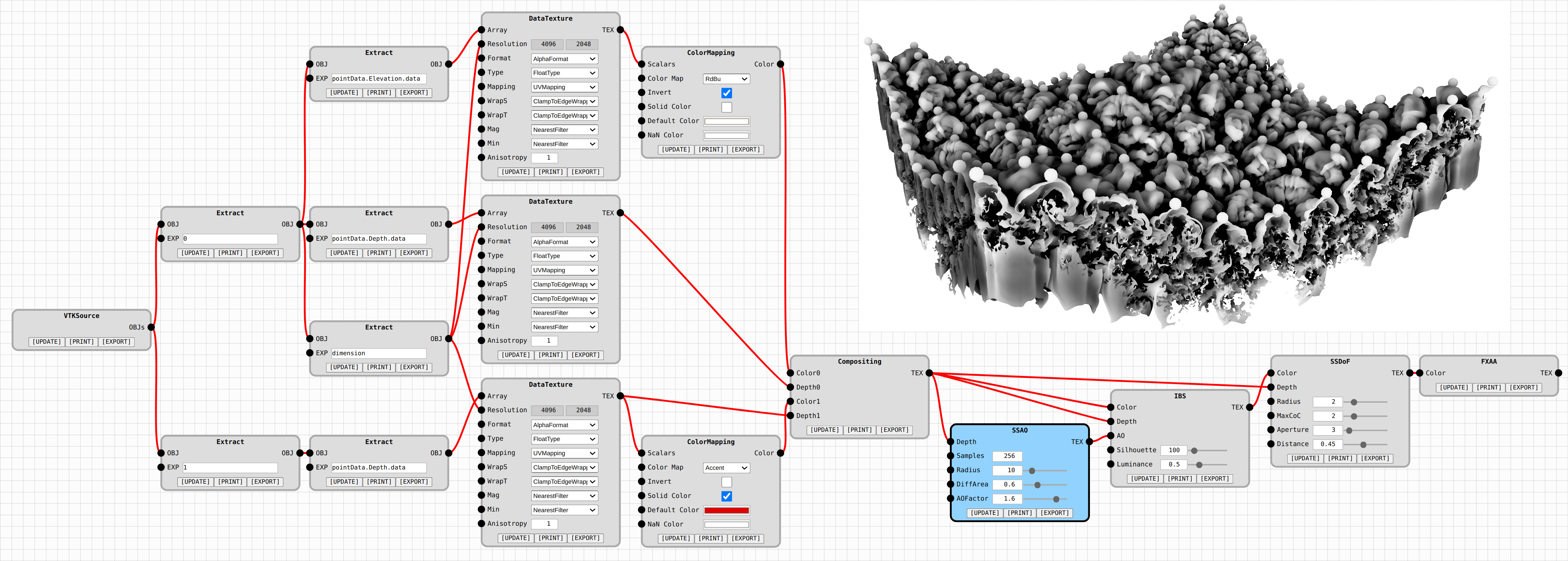}
    \vspace*{-2em}
    \caption{
        This figure shows the web-based deferred rendering front end that consists of a graph-based pipeline editor, and a render view that displays the output of a selected pipeline element.
        Currently the interface shows the output of the the screen space ambient occlusion (SSAO) filter (blue node).
        Every filter of the deferred rendering pipeline is shown as a node, where the filter's input and output parameters are displayed as ports (circles) on the left and right side of the corresponding node, respectively.
        Values of input parameters can be controlled with a widget shown on the right side of the corresponding input port.
        The connection between two ports (red lines) indicates that an output parameter is used as an input parameter of another filter.
        Users can conveniently create new nodes, connect them, and update their parameters while observing the effects on the resulting rendering at interactive framerates.
        \vspace*{-1.5em}
    }
    \label{fig_frontend}
\end{figure*}

With the continuous increase in computational power, the I/O bandwidth bottleneck becomes the limiting factor for interactive data analysis and visualization.
For instance, supercomputers are capable of simulating complex physical phenomena by performing $10^{18}$~floating operations per second, but bandwidth constraints still frequently prohibit storing every simulation state in its entirety for later exploration.
This problem gave rise to various techniques that aim to compute and store analysis and visualization products while the \edit{simulation} state still resides in machine memory.
In particular, Ahrens et al.~\cite{ahrens2015image} propose to sample in situ the parameter space (isovalues, timesteps, ...) and the visualization space (camera locations, clipping planes, ...) via images, and then store them in a so-called Cinema database.
Such a structured image database and can later be browsed post hoc along the sampling axes to emulate real-time visualization that is independent of the actual size and complexity of the depicted data.
However, the resulting post hoc visualization is limited to the images stored in the Cinema database, and this approach requires foresight about many visualization parameters that are usually adjusted interactively (such as suitable transfer functions~and~lighting~settings).

\newpage

The concept of deferred rendering~\cite{deering1988triangle,liang2000deferred} can overcome this issue by decoupling the information required for shading (e.g., depth buffers and scalar textures) from the shading itself (e.g., color mapping and lighting).
In the context of in situ visualization, this concept can be used to only store the underlying shading information in the Cinema database, and postpone final shading to the post hoc exploration phase, at which point visualization parameters that were not known a priori can be interactively adjusted.

This paper describes Cinema Darkroom (CD): an open-source deferred rendering framework that enables high-fidelity post hoc visualization~(\autoref{fig_teaser}).
CD follows the analogy of analog photography, where a scene is first captured on film and later developed in a darkroom; hence the name.
CD consists of two components: a c++ module implemented in the Topology ToolKit~(TTK)~\cite{ttk2017} that stores in situ the underlying shading information---collectively referred to as geometry buffers~\mbox{(G-Buffers)}---in a Cinema database, and a web-based deferred rendering front end that enables analysts to interactively retrieve, compose, and shade G-Buffers through a flexible shading network~(\autoref{fig_frontend}).
This network consists of individual, interconnectable shading passes that feature various image-based rendering techniques (\autoref{sec_cd}).
As demonstrated in \autoref{sec_results}, CD is versatile, extendable, and compatible with existing tool chains designed for in situ visualization and analysis.
To summarize, we make the following contributions:
\begin{itemize}
  \setlength\itemsep{-0.2em}

  \item \edit{We apply deferred rendering in the context of in situ visualization, where geometry buffers are stored in situ, and deferred shading is performed during post hoc exploration.}
  \item \edit{We describe the design and implementation of a flexible shading system that features a node-based workflow.}
  \item \edit{We illustrate the capabilities of CD via a variety of real-world examples, and show that image-based shading solely based on G-Buffers can be used for high-fidelity visualization.}

\end{itemize}

\section{Related Work}
Ahrens et al.~\cite{ahrens2015image} proposed Cinema: the first image-based approach to extreme-scale in situ visualization and analysis.
Their research is based on the rational that no matter how big the depicted dataset is, in the end visualization happens on a viewport with a fixed number of pixels.
So to circumvent the bandwidth bottleneck, they propose to store, at simulation runtime, visualizations instead of the depicted datasets, since the visualizations are several magnitudes smaller than the data they are derived from.
Specifically, they propose to sample the parameter space (isovalues, timesteps, etc.) and the visualization space (clipping planes, color maps, camera angles, etc.), and then generate for each sample a visualization that is uniquely identified by the sample values.
This association enables analysts to later browse the resulting image database along the sampling axes, which emulates real-time data visualization of any size.
\edit{Moreover, advanced Cinema front ends~\cite{dntg2020,aldrich2019query} can use feature-centric queries to retrieve corresponding images.}\vspace*{-0.045em}

However, Cinema requires foresight about suitable samplings, and post hoc visualization is limited to the elements in the database.
\edit{To address these issues, O’Leary et al.~\cite{o2016cinema} extended the initial concept of Cinema databases by instead of storing final visualizations, they propose to store geometry buffers~(G-Buffers)~\cite{deering1988triangle,liang2000deferred} that record general information of each pixel, such as the distance to the camera and the depicted scalar value.
This approach enables post hoc color mapping, compositing, and shading, which not only increases flexibility during post hoc exploration, but also decreases the size of the image database.
This concept of storing and processing G-Buffers is formalized in the latest Cinema specification~\cite{rogers2018cinemaspec}.}\vspace*{-0.045em}

We describe a system that fully utilizes G-Buffers stored in a Cinema database for high-fidelity post hoc visualization.
In contrast to related work aimed at facilitating post hoc exploration on a conceptual level
---such as the \emph{Explorable Images} approach~\cite{tikhonova2010explorable}, \emph{Contour Tree Depth Images}~\cite{biedert2015}, or \emph{Space Time Volumetric Depth Images}~\cite{fernandes2014space}---
we aim to investigate how a flexible post hoc shading system can maximize the value of G-Buffer data stored in an in situ generated Cinema database.
At heart, our system is based on a deferred rendering approach~\cite{liang2000deferred,deering1988triangle} that feeds selected G-Buffers into a flexible shading architecture consisting of individual, interconnectable shading passes~(\autoref{fig_frontend}).
Such shading networks are commonly used in production tools like \emph{Blender}~\cite{blender2018} and \emph{Autodesk 3D Studio Max}~\cite{harper2012mastering}, or in game engines such as the \emph{Unity Engine}~\cite{unityengine} and the \emph{Unreal Engine}~\cite{unrealengine}.
To this end, CD features various deferred rendering-based shading techniques~\cite{rogers2018cinemaspec,ritschel2009approximating,luft2006image,mcintosh2012efficiently,dntg2020,lottes2009fast}~(\autoref{sec_frontend}).
Moreover, due to its file system-based API, CD is compatible with existing frameworks designed for in situ database generation, such as \emph{ParaView Catalyst}~\cite{ayachit2015paraview} and \emph{Alpine Ascent}~\cite{larsen2017alpine}, as well as any other framework capable of writing out G-Buffers as files.

\newcommand{\scale}{4.1cm}
\newcommand{\scaleX}{4.15}
\newcommand{\scaleY}{1}

\newcommand{\legend}[2]{
    \setlength{\fboxrule}{0.25pt}

    \node at (#1,#2+0.5*\scale){60};
    \node at (#1,#2){30};
    \node at (#1,#2-0.5*\scale){ 0};
    \node at (#1,#2-0.57*\scale){Density};

    \draw (#1-0.04,#2+0.5*\scale) -- ++(-0.05,0);
    \draw (#1-0.04,#2)            -- ++(-0.05,0);
    \draw (#1-0.04,#2-0.5*\scale) -- ++(-0.05,0);
    \draw (#1+0.04,#2+0.5*\scale) -- ++(+0.05,0);
    \draw (#1+0.04,#2)            -- ++(+0.05,0);
    \draw (#1+0.04,#2-0.5*\scale) -- ++(+0.05,0);

    \node at (#1-0.1,#2) {
        \fbox{\includegraphics[width=1em,height=\scale]{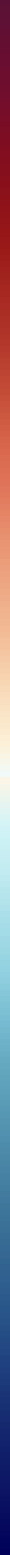}}
    };

    \node at (#1+0.1,#2) {
        \fbox{\includegraphics[width=1em,height=\scale]{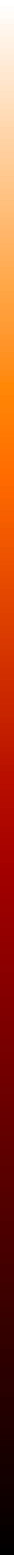}}
    };
}

\newcommand{\subfigure}[4]{
    \setlength{\fboxsep}{0pt}
    \node at (#1,#2) {
        \includegraphics[height=\scale]{#3}

    };

    \node at (#1,#2-0.57*\scale) {#4};

}

\begin{figure*}

    \begin{center}
        \begin{tikzpicture}[xscale=\scaleX,yscale=\scaleY]

            \subfigure{-1.7}{0}{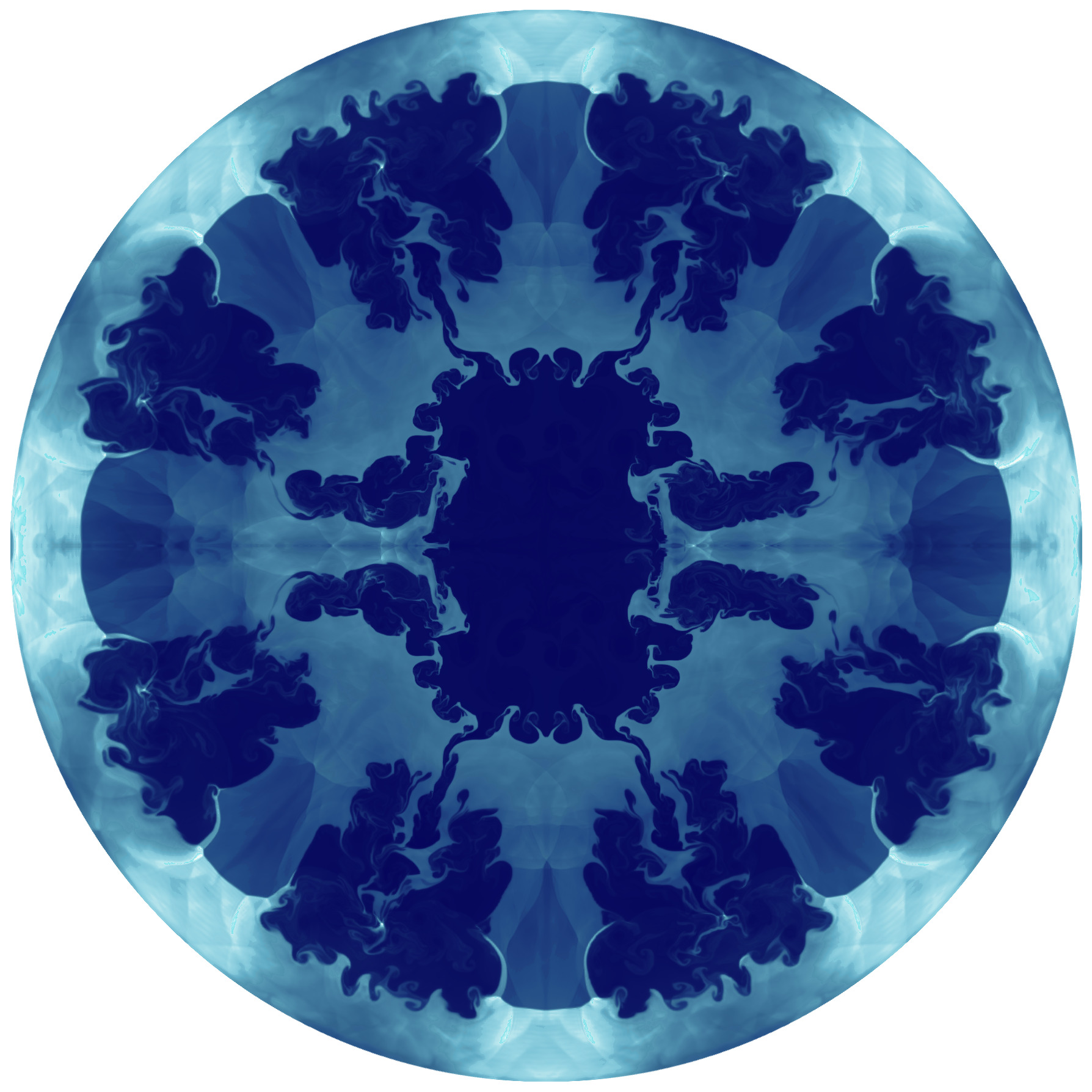}{cycle 75,000};
            \subfigure{-0.7}{0}{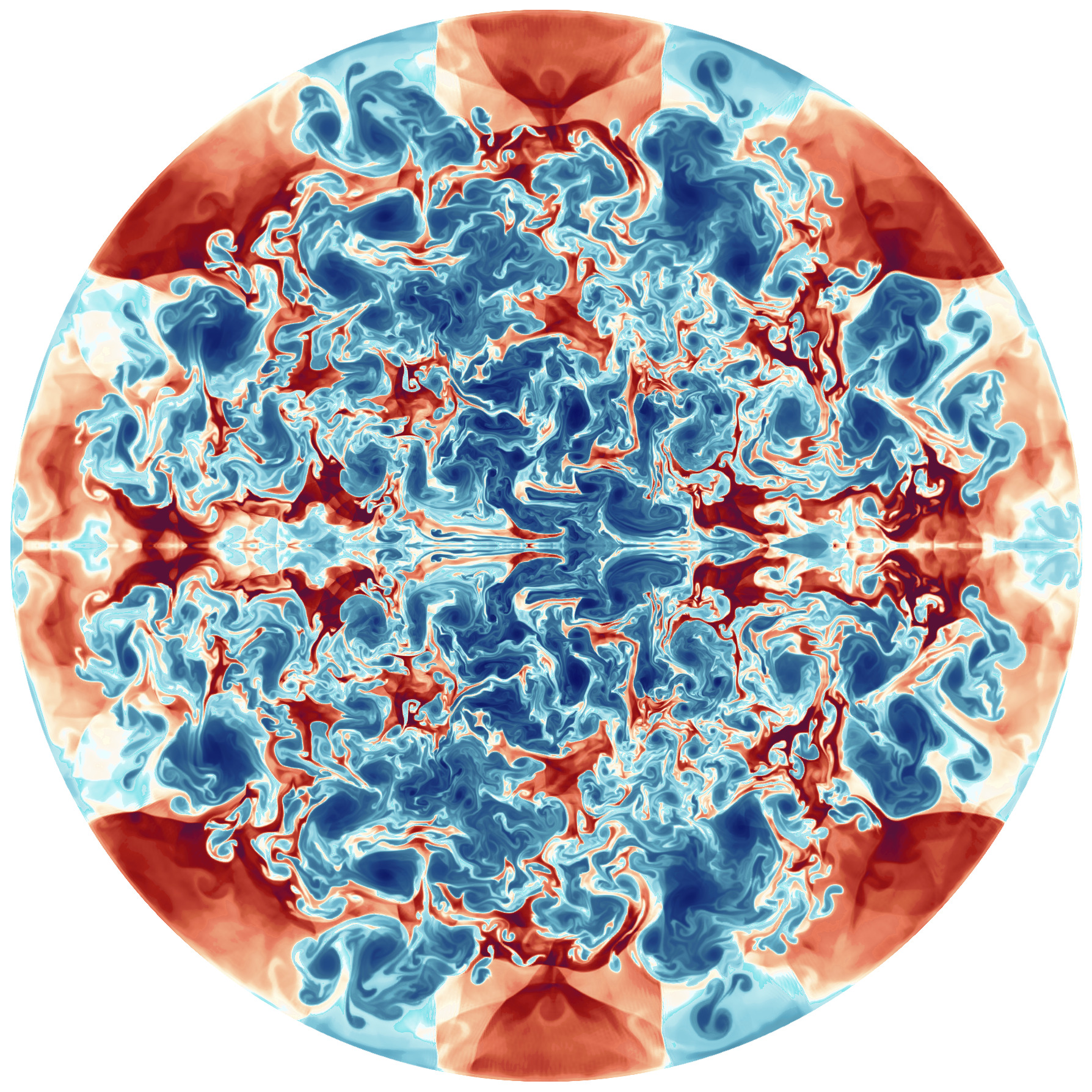}{cycle 134,220};

            \legend{0}{0};

            \subfigure{0.7}{0}{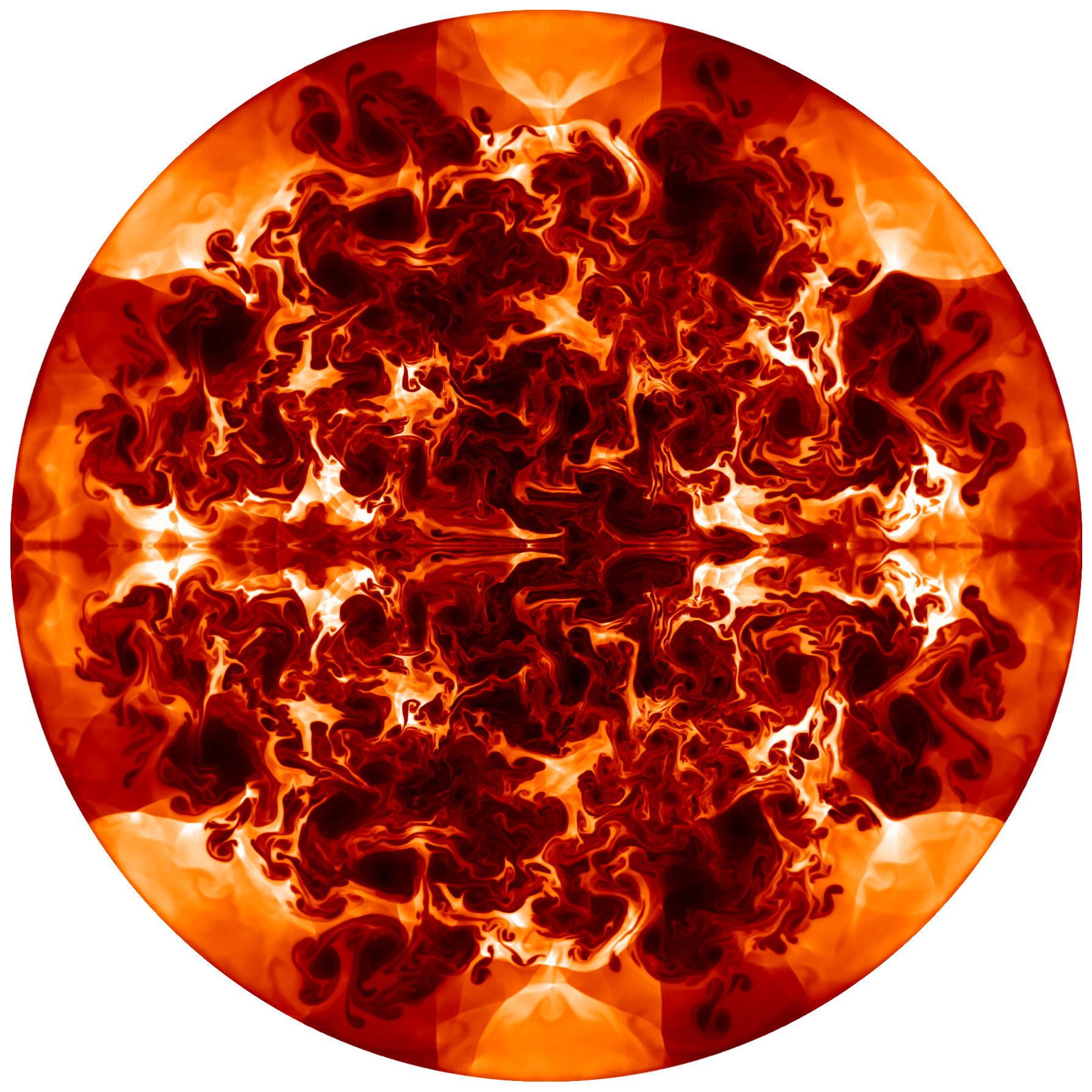}{cycle 134,220};
            \subfigure{1.7}{0}{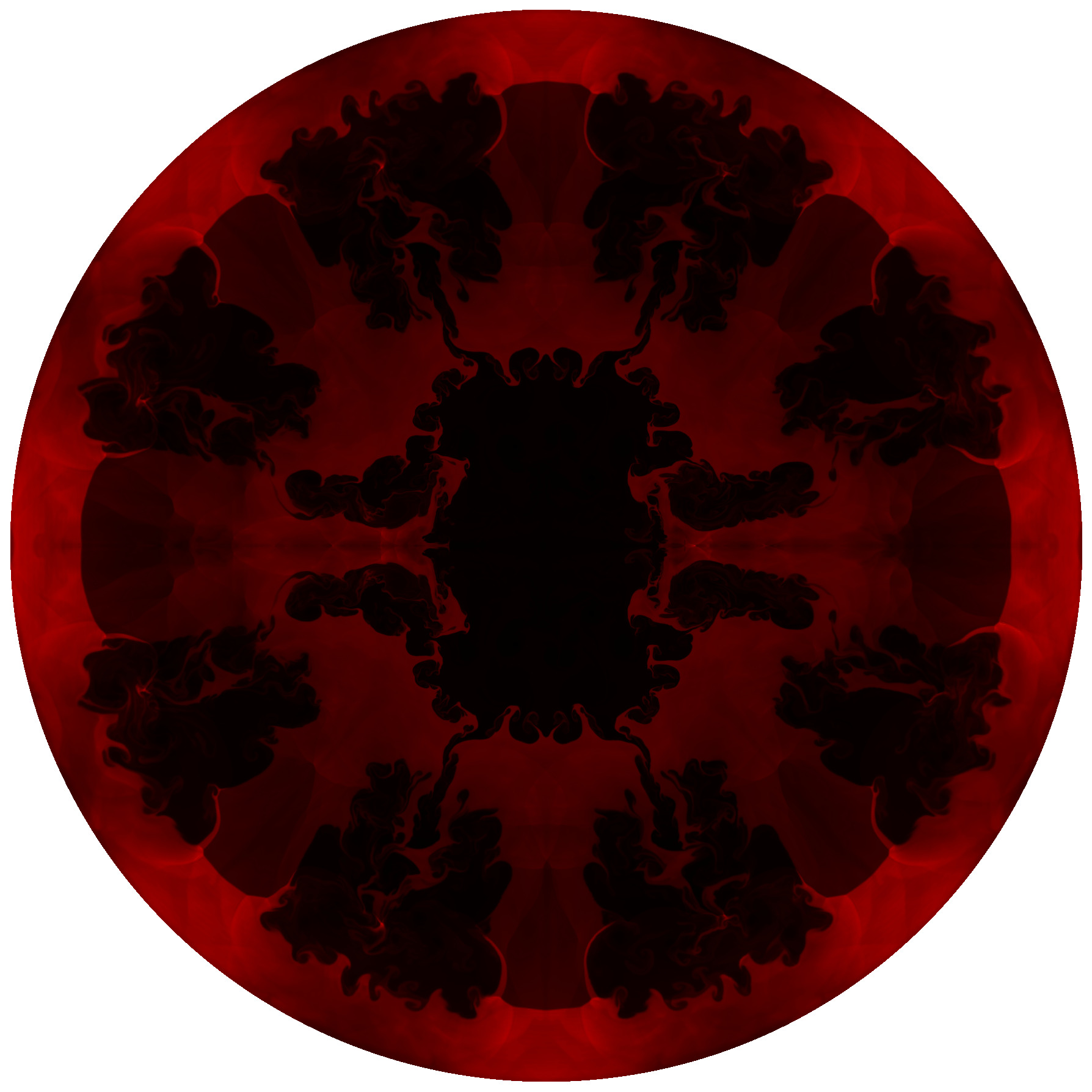}{cycle 75,000};

        \end{tikzpicture}%
    \end{center}
    \vspace*{-2.5em}

    \caption{
        This case study demonstrates the simplest use case of CD where it is only necessary to apply post hoc color mapping.
        The images show two simulation cycles of an idealized, two material, inertial confinement fusion implosion test problem in 2D, simulated using Blast~\cite{blast}.
        A full description of the problem, including initial conditions, are described by Bello-Maldonado et al.~\cite{bello2020matrix}.
        During the course of the simulation, researchers at the Lawrence Livermore National Laboratory created a highly time-resolved scalar image database---consisting of over 2 thousand $2048^2$ pixel images---of the density field using Ascent~\cite{larsen2017alpine} and Devil Ray\cite{DevilRay}, which are released under LLNL-PHOTO-812321.
        \edit{Excluding further compression benefits, the resulting image database is roughly $20\times$ smaller than storing the original unstructured quad mesh.
        As CD uses a filesystem-based interface to read G-Buffers, one can feed the stored image database directly into the deferred rendering pipeline to explore~suitable~colormaps}.
        \vspace*{-1em}
    }
    \label{fig_ifc}
\end{figure*}

\section{Cinema Darkroom (CD)}
\label{sec_cd}

CD consists of a C++ module that handles G-Buffer generation, \edit{and a JavaScript-based deferred rendering front end for post hoc production visualization in the browser.}

\subsection{G-Buffer Database Generation}
    To generate G-Buffers, CD provides a new Visualization ToolKit~(VTK)~\cite{vtkbook} filter called \textit{CinemaImaging} that requires two inputs:
    1) a dataset that is going to be depicted in form of a triangulation, and
    2) a sampling grid storing camera locations and calibrations in form of a point cloud.
    \edit{The sampling grid can be specified manually~\cite{ahrens2015image} or automatically~\cite{lukasczyk2018voidga} based on the depicted dataset.}
    The filter then generates geometry buffers for each sample point via an Embree-based raytracer~\cite{embree2014}.
    By default, the filter stores in the G-Buffer the depth buffer (the distance to the camera), and a scalar buffer for each scalar field defined on the input dataset.
    As shown in \autoref{sec_results}, shading based on the depth buffers alone already provides good results, as 3D-world positions and surface normals can be accurately reconstructed in image-space, since the exact camera calibrations are known~\cite{lukasczyk2018voidga}.
    However, if desired, it is also possible to explicitly store position and normal vectors in the \mbox{G-Buffers}.
    Moreover, \edit{depending on the desired post hoc rendering effects}, one can store additional information in the \mbox{G-Buffers}, such as true global illumination or motion blur.

    Each resulting \mbox{G-Buffer} is stored in a Cinema database~\cite{rogers2018cinemaspec} (e.g., in \textit{vtkImageData} format), which can be accessed by the web-based front end.
    Thus, every framework capable of writing out \mbox{G-Buffers} as files is compatible with the proposed workflow.
    For example, the \mbox{G-Buffers} used in \autoref{fig_ifc} were stored in situ via a highly parallel raytracer~\cite{moreland2016vtk} and the Ascent in situ framework~\cite{larsen2017alpine}.

\subsection{Deferred Rendering Front End}
\label{sec_frontend}

\autoref{fig_frontend} shows the web-based deferred rendering front end---written solely in client-side JavaScript and html---that consists of a graph-based pipeline editor and a render view that shows the output of a selected filter.
The interface is modeled after established tools such as the material editor of Blender~\cite{blender2018} and the pipeline editor of Inviwo~\cite{jonsson2019inviwo}.
Specifically, every filter used in the current pipeline is shown as a node, where its input and output parameters are shown as ports on the left and right side of the node, respectively.
Additionally, the value of an input parameter can be controlled via a widget that is displayed on the right side of the port.
If a filter requires the execution of shader code, then its input parameters are automatically passed to the vertex and fragment shaders as uniforms.
The actual rendering is performed via Three.js~\cite{dirksen2013learning} and WebGL~\cite{marrin2011webgl}, \edit{where input and output G-Buffers are stored as Framebuffer Objects}.
An output of one filter can be used as input of another filter, where the pipeline follows a push model.
Thus, as soon as a new input value for a filter is available, the filter updates its output, which then triggers an update of filters that use the updated output as an input, and so forth.
Output ports can be connected to input ports via drag an drop, where established connections are shown via red edges.

\renewcommand{\scaleX}{2.8}
\renewcommand{\scaleY}{1}

\renewcommand{\subfigure}[4]{
    \setlength{\fboxsep}{0pt}
    \node[rotate=90] at (#1-0.07,#2) {
        \includegraphics[width=0.265\textwidth]{#3}
    };
    \node at (#1,#2-2.7) {#4};
}

\begin{figure}

    \begin{center}
        \hspace*{-30em}%
        \begin{tikzpicture}[xscale=\scaleX,yscale=\scaleY]
            \subfigure{-1}{0}{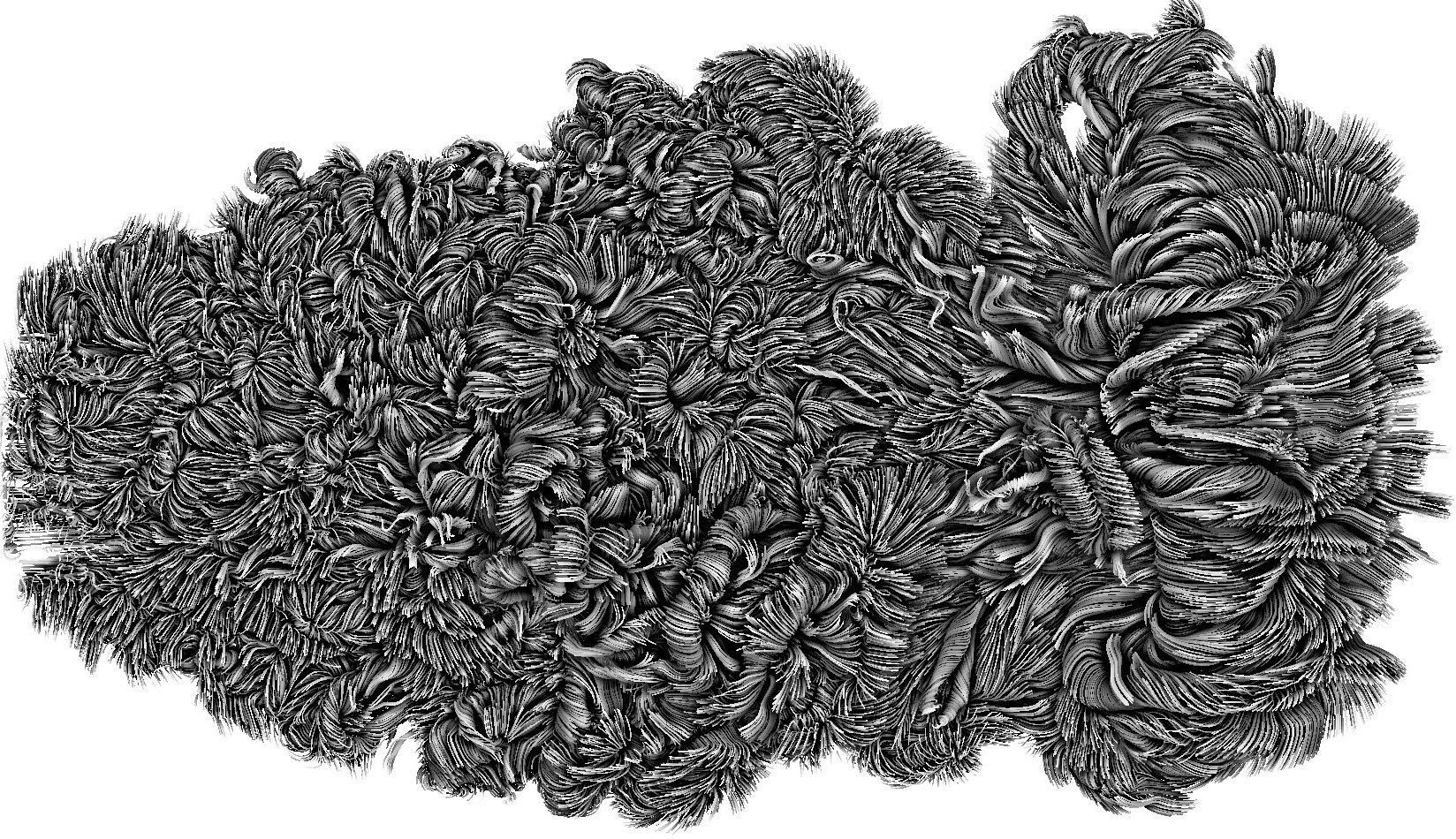}{SSAO Radius $0.3\%$};
            \subfigure{0}{0}{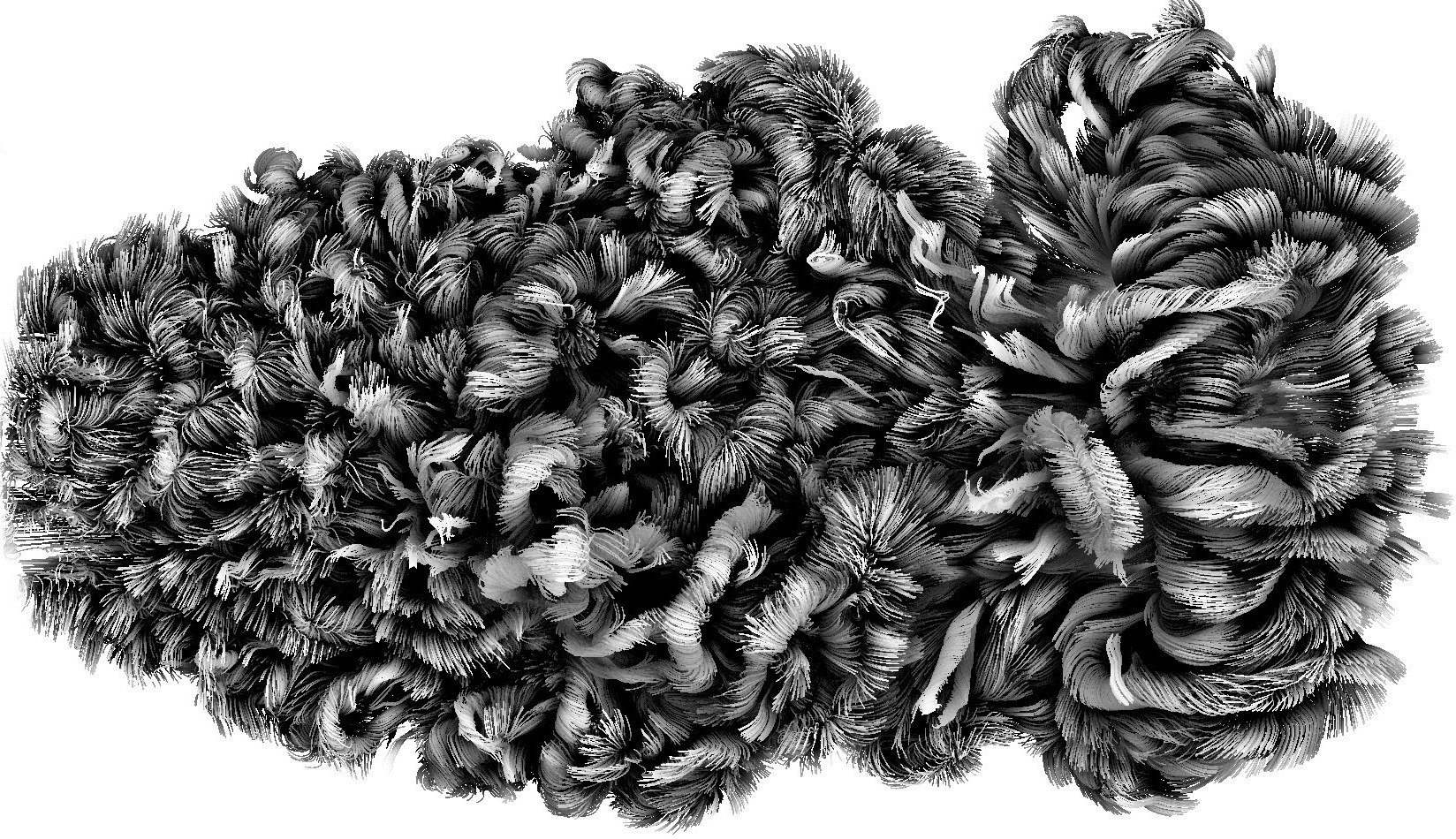}{$1\%$};
            \subfigure{1}{0}{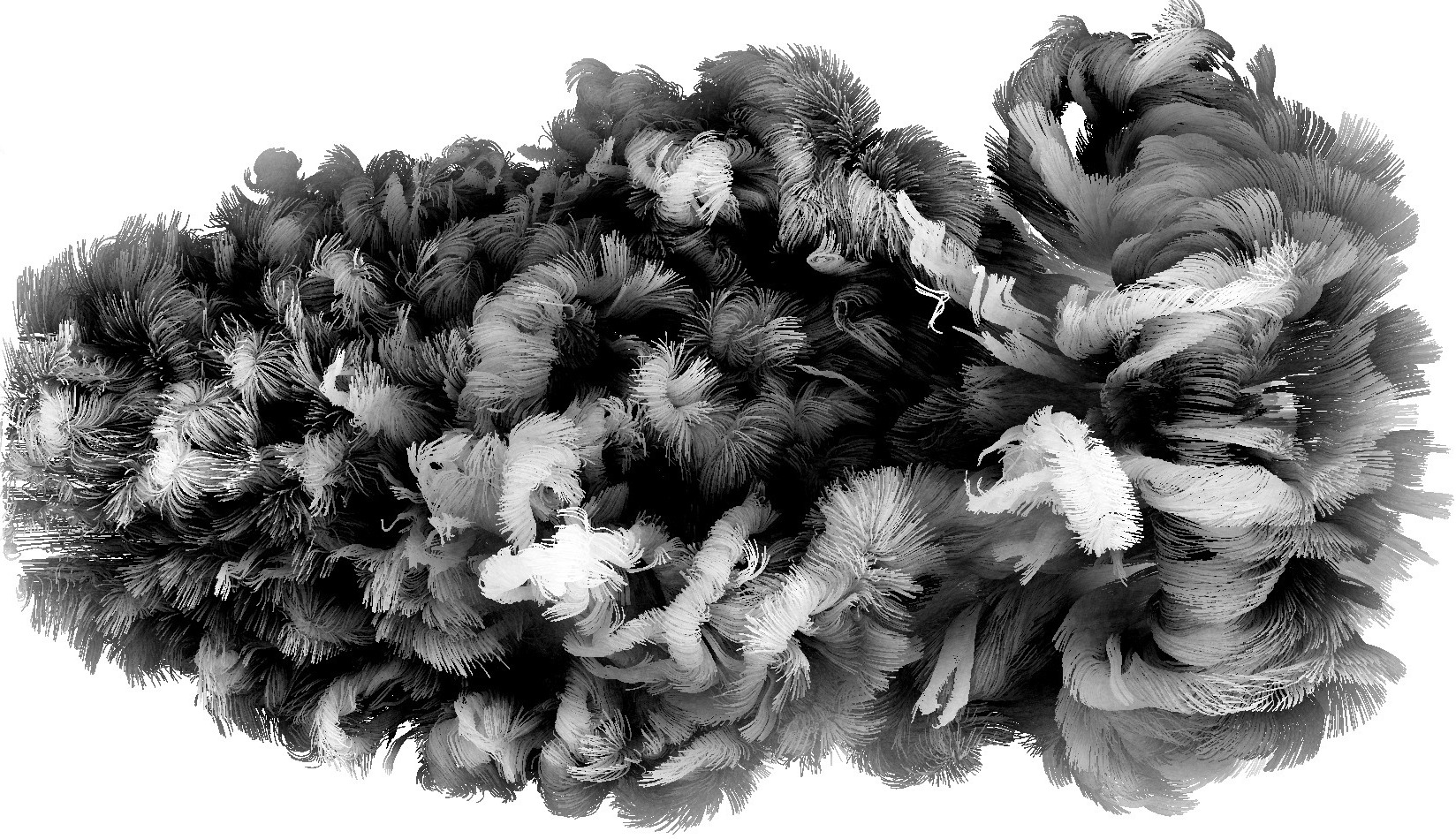}{$3\%$};
        \end{tikzpicture}%
        \hspace*{-30em}\vspace*{-2em}
    \end{center}

    \caption{
        Impact of the SSAO sampling radius (specified in percent of the image height) on the approximation of global illumination in the Jet dataset~(\autoref{fig_jet}).
        Small radii are better to contrast local neighborhoods, while large radii can approximate global lighting.
        CD can combine SSAO at different scales to profit from both advantages~(\autoref{fig_jet} right).
    }
    \label{fig_ao}
\end{figure}

Besides multipurpose filters used for data transformations, CD currently features the following image-based rendering techniques:\vspace{-0.25em}
    \begin{itemize}[leftmargin=8\parindent]
        \setlength\itemsep{-0.2em}
        \item[\textbf{ColorMapping}] Post Hoc Color Mapping~\cite{rogers2018cinemaspec};
        \item[\textbf{Compositing}] Depth Image Based Compositing~\cite{dntg2020};
        \item[\textbf{SSAO}] Screen Space Ambient Occlusion~\cite{ritschel2009approximating};
        \item[\textbf{SSDD}] Screen Space Depth Darkening~\cite{luft2006image};
        \item[\textbf{SSDoF}] Screen Space Depth of Field~\cite{mcintosh2012efficiently};
        \item[\textbf{IBS}] Image-Based Silhouettes~\cite{dntg2020}; and
        \item[\textbf{FXAA}] Fast Approximate Anti-Aliasing~\cite{lottes2009fast}.\vspace{-0.25em}
    \end{itemize}
\edit{New filters can be added to CD with minimal overhead.
To add a filter, developers just needs to 1)~inherit form the abstract filter class, 2)~define input and output parameters, and 3)~provide the code that transforms input to output parameters (e.g., shader code).}

\delete{CD also features an export mechanism that generates human readable JavaScript code that will recreate the current configured pipeline; including all parameter settings and connections.
This feature makes it possible to code pipelines instead of configuring them manually.}

\renewcommand{\scaleX}{2.9}
\renewcommand{\scaleY}{1}

\renewcommand{\subfigure}[4]{
    \setlength{\fboxsep}{0pt}
    \node[rotate=90] at (#1,#2) {
        \includegraphics[width=0.33\textwidth]{#3}
    };
    \node at (#1,#2-2.7) {#4};
}

\renewcommand{\legend}[5]{
    \setlength{\fboxrule}{0.25pt}

    \node at (0,#1+0.3) {#2};
    \node[anchor=east] at (-0.95,#1) {#4};
    \node[anchor=east] at ( 1.2,#1) {#5};
    \node at (0,#1) {
        \fbox{\includegraphics[width=0.3\textwidth,height=1em]{#3}}
    };

}

\begin{figure}

    \begin{center}
        \hspace*{-30em}%
        \begin{tikzpicture}[xscale=\scaleX,yscale=\scaleY]
            \subfigure{-1}{0}{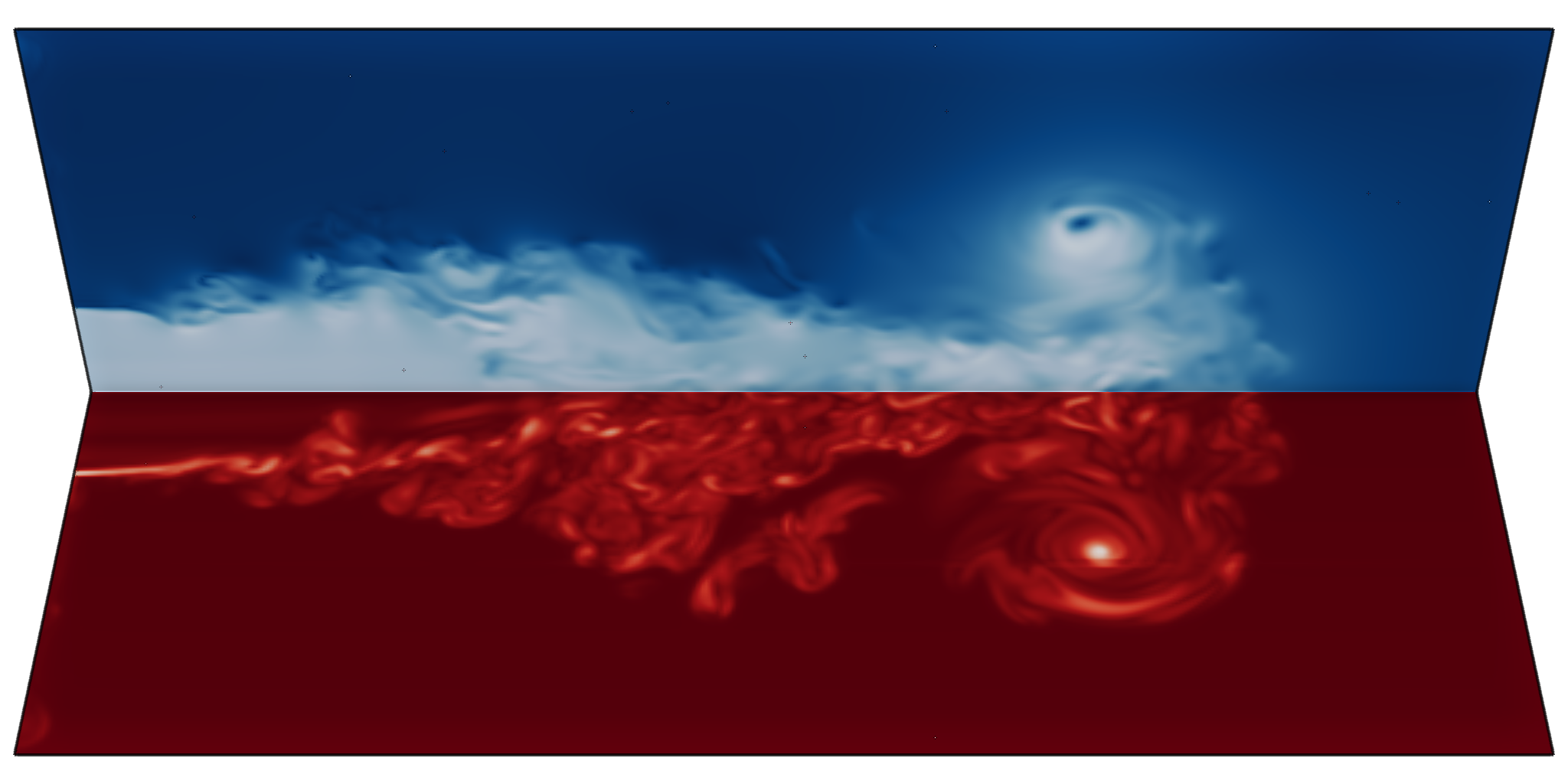}{};
            \subfigure{0}{0}{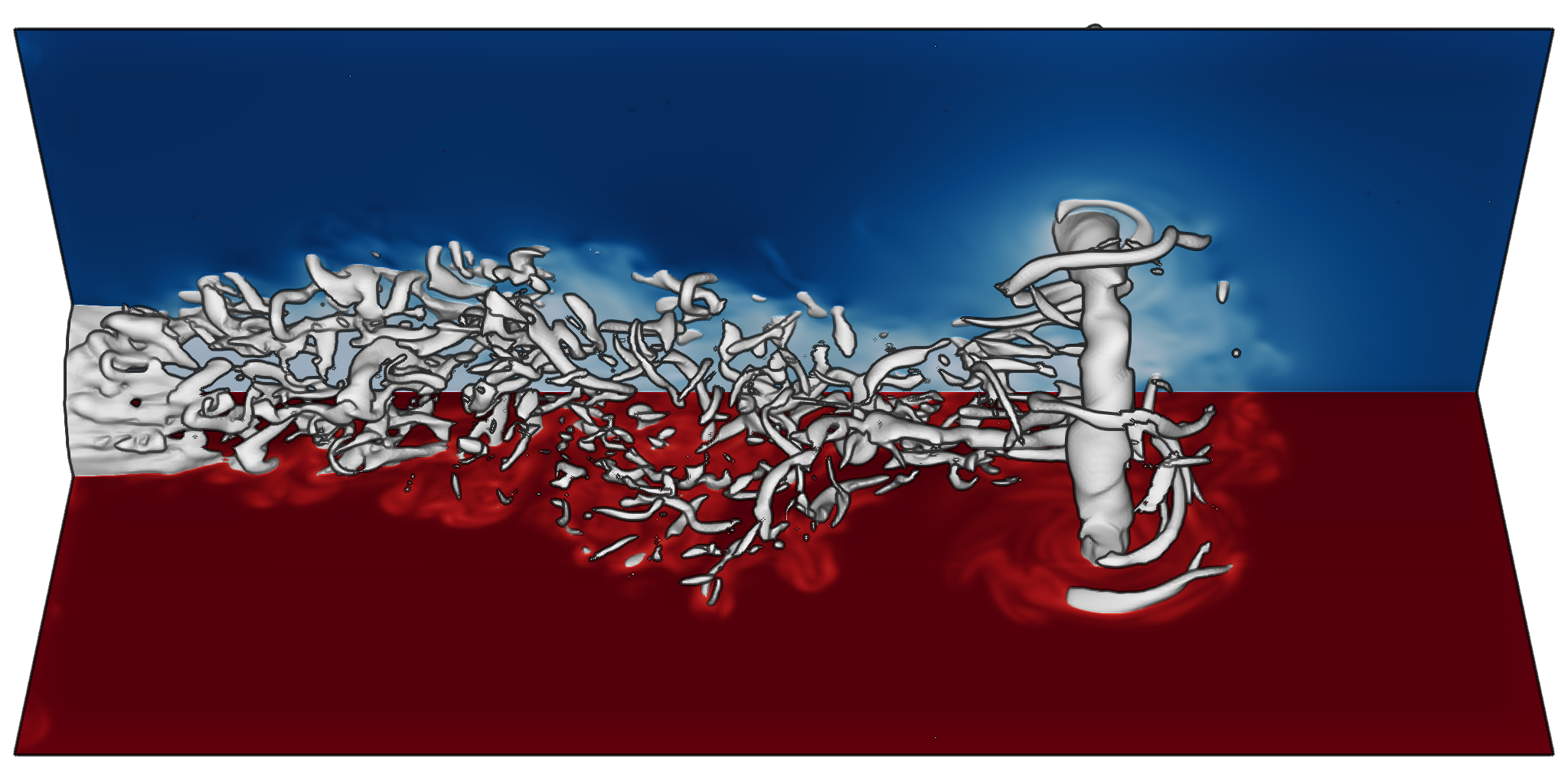}{};
            \subfigure{1}{0}{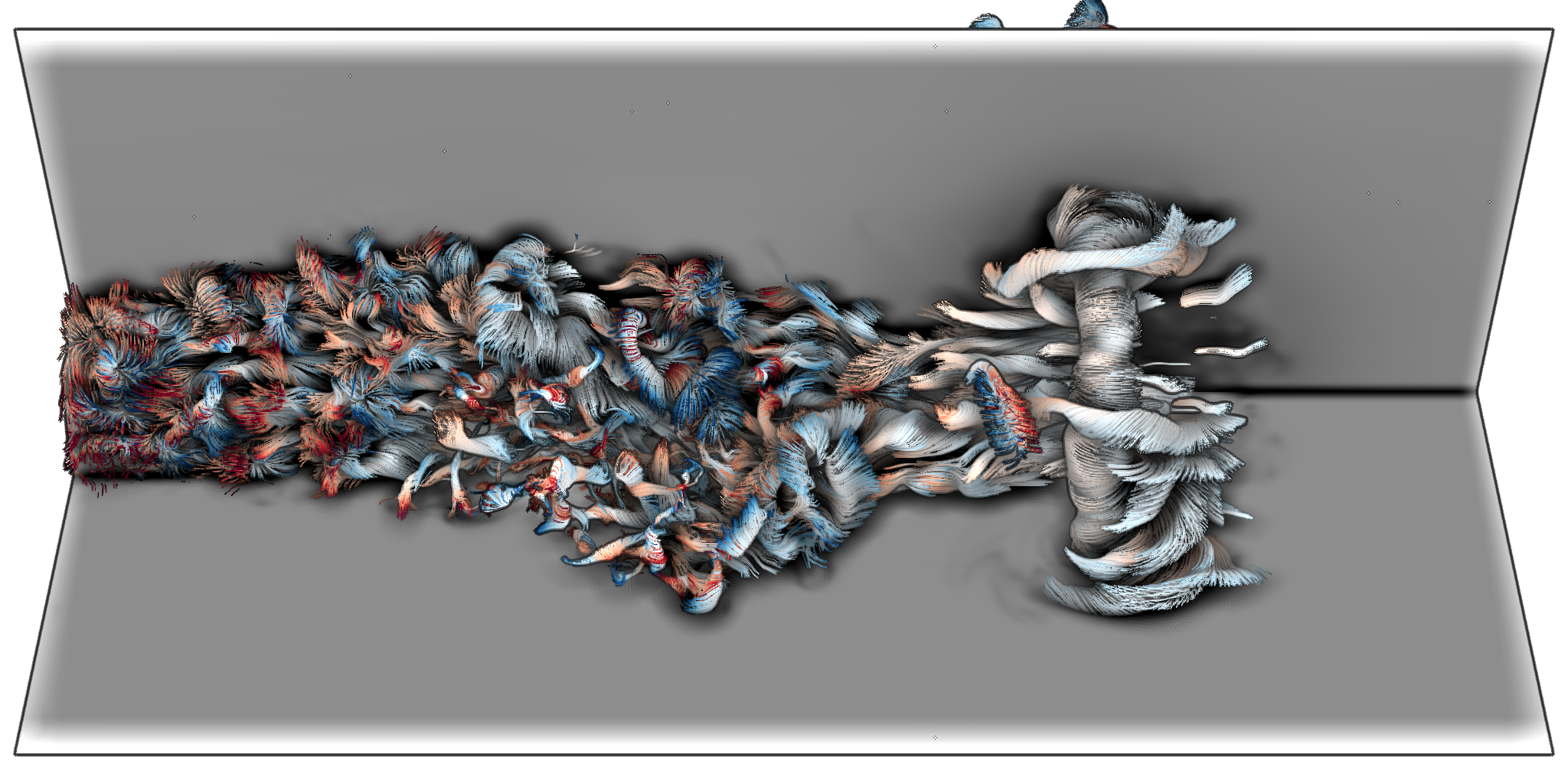}{};

            \legend{-3.6}{Velocity Magnitude (Left Cutting Plane)}{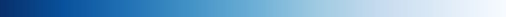}{0}{14};
            \legend{-4.4}{Vorticity Magnitude (Right Cutting Plane)}{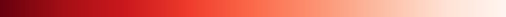}{0}{1700};
            \legend{-5.2}{Integration Time (Streamlines)}{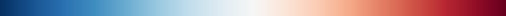}{-1}{1};
        \end{tikzpicture}%
        \hspace*{-30em}\vspace*{-2em}
    \end{center}

    \caption{
        The jet dataset~\cite{garthJetData} is a \emph{Gerris Flow Solver}~\cite{popinet2003gerris} simulation of a fluid jet entering a medium at rest.
        Vorticity and velocity magnitudes are computed from the original flow field, which is visualized via cutting planes (left), a vorticity isosurface (center), and streamlines seeded on the isosurface (right).
        Instead of storing in situ the Cartesian product of all possible visualization elements, CD can dynamically combine any set of visualization elements post hoc via depth compositing.\vspace*{-1em}}
    \label{fig_jet}
\end{figure}

\section{Results}
\label{sec_results}

This section demonstrates on several real-world datasets the versatility and flexibility of CD, where the context of the presented visualizations are reported in the captions of Figures~3-7.

It is important to note that all presented visualizations are solely shaded based on depth buffers (to approximate lighting) and scalar images (for color mapping).
So no actual lighting information is stored in the corresponding Cinema databases.
Thus, the most significant filter that impacts visual quality is Screen Space Ambient Occlusion~(SSAO)~\cite{ritschel2009approximating} (or alternatively Screen Space Depth Darkening~(SSDD)~\cite{luft2006image}) as it approximates global illumination.
Input parameters of this filter control the sample number and the sampling radius to approximate the local neighborhood of each pixel, where small radii tend to produce sharp shadows, while large radii produce soft shadows~(\autoref{fig_ao}).
The number of samples then controls the quality of the approximation.
The pipeline editor makes it possible to interactively search for suitable parameters, and to combine multiple SSAO filters to approximate shadows at different scales~(\autoref{fig_jet}).

The SSAO technique implemented in CD works well in all presented case studies, except for the particle visualization shown in~\autoref{fig_part}, since the current implementation is designed to approximate shadows on surface-like structures.
This technique still produces good results for the visualization of streamlines~(Figs.~\ref{fig_ao}-\ref{fig_jet}) and jagged surfaces~(\autoref{fig_stone}), but causes visual artifacts for highly discontinuous depth buffers which are prominent in particle visualizations.
This limitation can be addressed by adding SSAO techniques that are optimized for such visualizations~\cite{engelke2019autonomous,grottel2014megamol,eichelbaum2013pointao}.

\renewcommand{\scale}{0.48\textwidth}
\renewcommand{\scaleX}{1}
\renewcommand{\scaleY}{2.9}

\renewcommand{\subfigure}[4]{
    \node at (#1,#2) {
        \includegraphics[width=\scale]{#3}
    };

}

\begin{figure}
    \begin{center}
        \begin{tikzpicture}[xscale=\scaleX,yscale=\scaleY]
            \subfigure{0}{0}{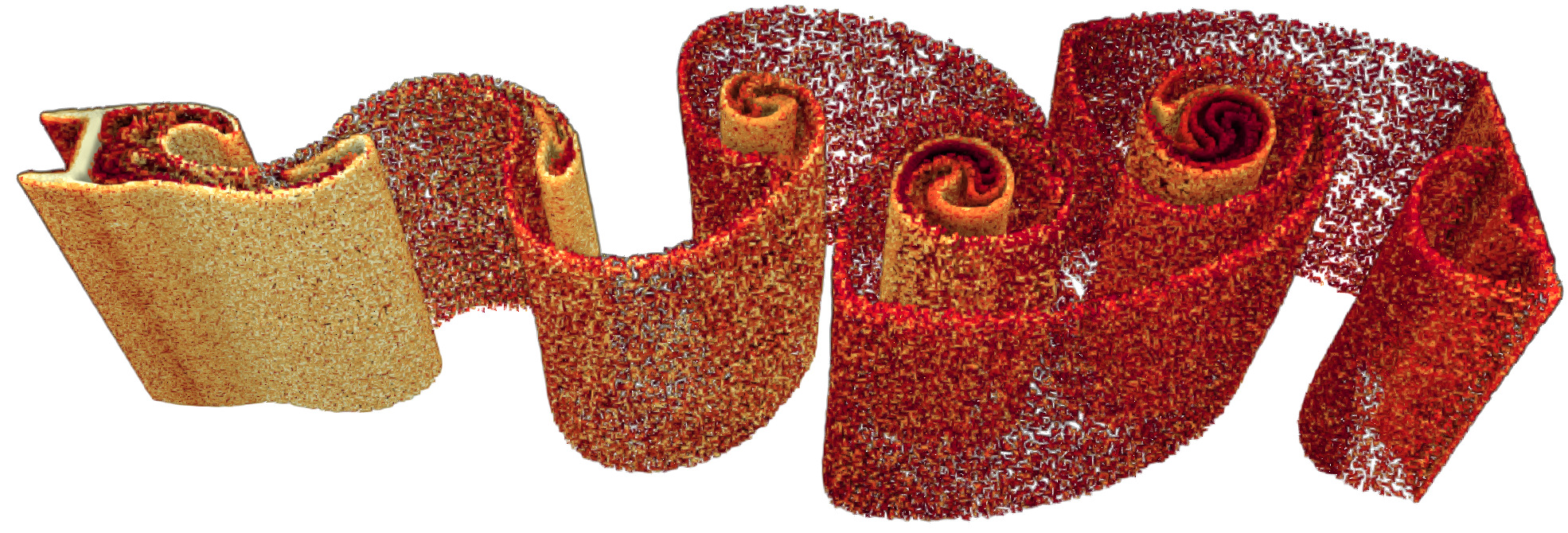}{cycle X};
            \subfigure{0}{-1}{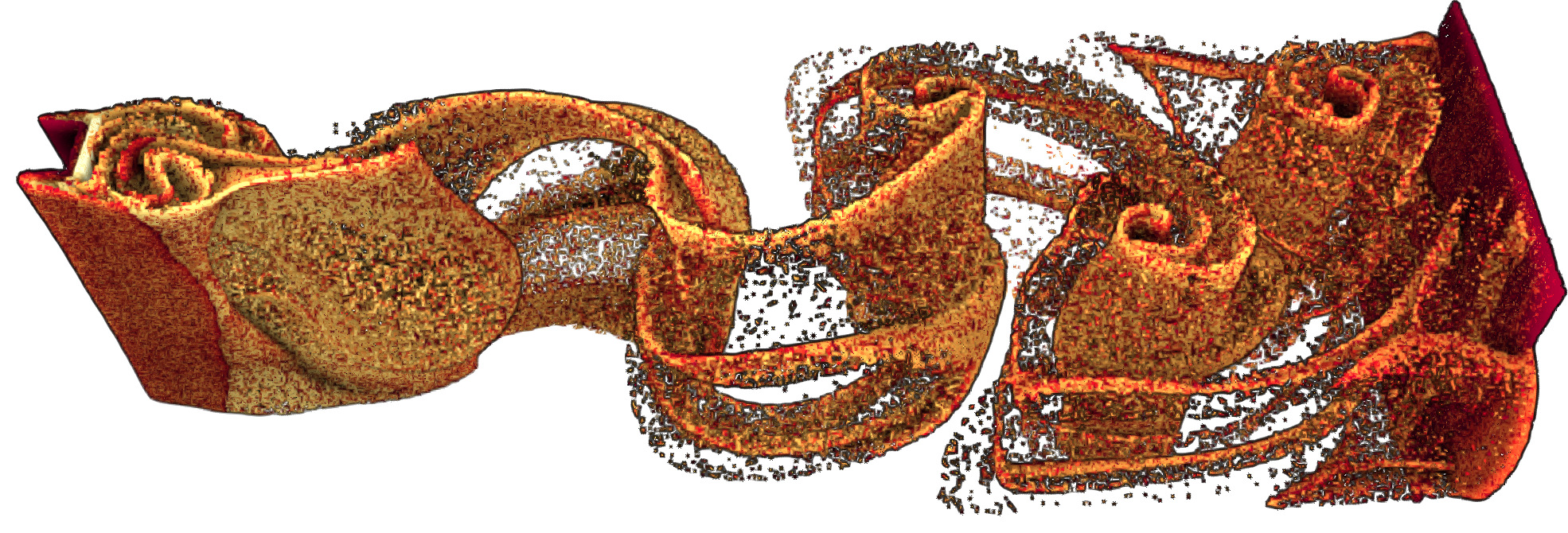}{cycle X};
        \end{tikzpicture}%
    \end{center}
    \vspace*{-2.5em}

    \caption{
        This case study uses a particle-based visualization to highlight vortex regions in a flow field around an obstacle located on the left side of the domain~\cite{popinet2003gerris,weinkauf10c}.
        Particles are continuously seeded in close proximity to the obstacle over the complete time span, where each timestep simulates roughly \edit{4 million particles}.
        Each particle is advected along its path line; creating streak surface like structures.
        The images show two different timesteps where particles are colored by seed time (from brown to red), and are shaded by SSAO and IBS.
        The current SSAO implementation in CD is optimized for surfaces, and produces visible artifacts for highly variant depth buffer neighborhoods, which are prominent in particle simulations.
        In the future, this problem can be addressed by supporting at least one of the SSAO algorithms explicitly developed for particle rendering~\cite{engelke2019autonomous,grottel2014megamol,eichelbaum2013pointao}.
    }
    \label{fig_part}
\end{figure}

\section{Conclusion}

This paper described Cinema Darkroom (CD): an interactive deferred rendering framework for high-fidelity post hoc visualization.
As demonstrated on \edit{several} real-world examples~(\autoref{sec_results}), CD is versatile, flexible, and compatible with existing tool chains designed for in situ visualization and analysis.

However, CD is currently limited to G-Buffers that depict fully opaque geometry.
This notably excludes post hoc volume rendering (except in its most basic form of directly storing color images in situ).
To address this issue, other techniques such as the \emph{Explorable Images} approach~\cite{tikhonova2010explorable} or generative models~\cite{berger2018generative} need to be investigated if they
can be integrated in a deferred rendering pipeline.
Finally, CD is designed to be easily extendable and we plan to include in future work more image-based shading techniques such as Image-Space Line Integral Convolution~\cite{Huang2012}, Scalable Ambient Obscurance~\cite{mcguire2012scalable}, and SSAO optimized for particle visualization~\cite{engelke2019autonomous,grottel2014megamol,eichelbaum2013pointao}.

\renewcommand{\scaleX}{2.2}
\renewcommand{\scaleY}{1}

\renewcommand{\subfigure}[4]{
    \setlength{\fboxsep}{0pt}
    \node at (#1,#2) {
        \includegraphics[width=0.24\textwidth]{#3}
    };
    \node at (#1,#2-2.7) {#4};
}

\begin{figure}

    \begin{center}
        \hspace*{-30em}%
        \begin{tikzpicture}[xscale=\scaleX,yscale=\scaleY]
            \subfigure{-1}{0}{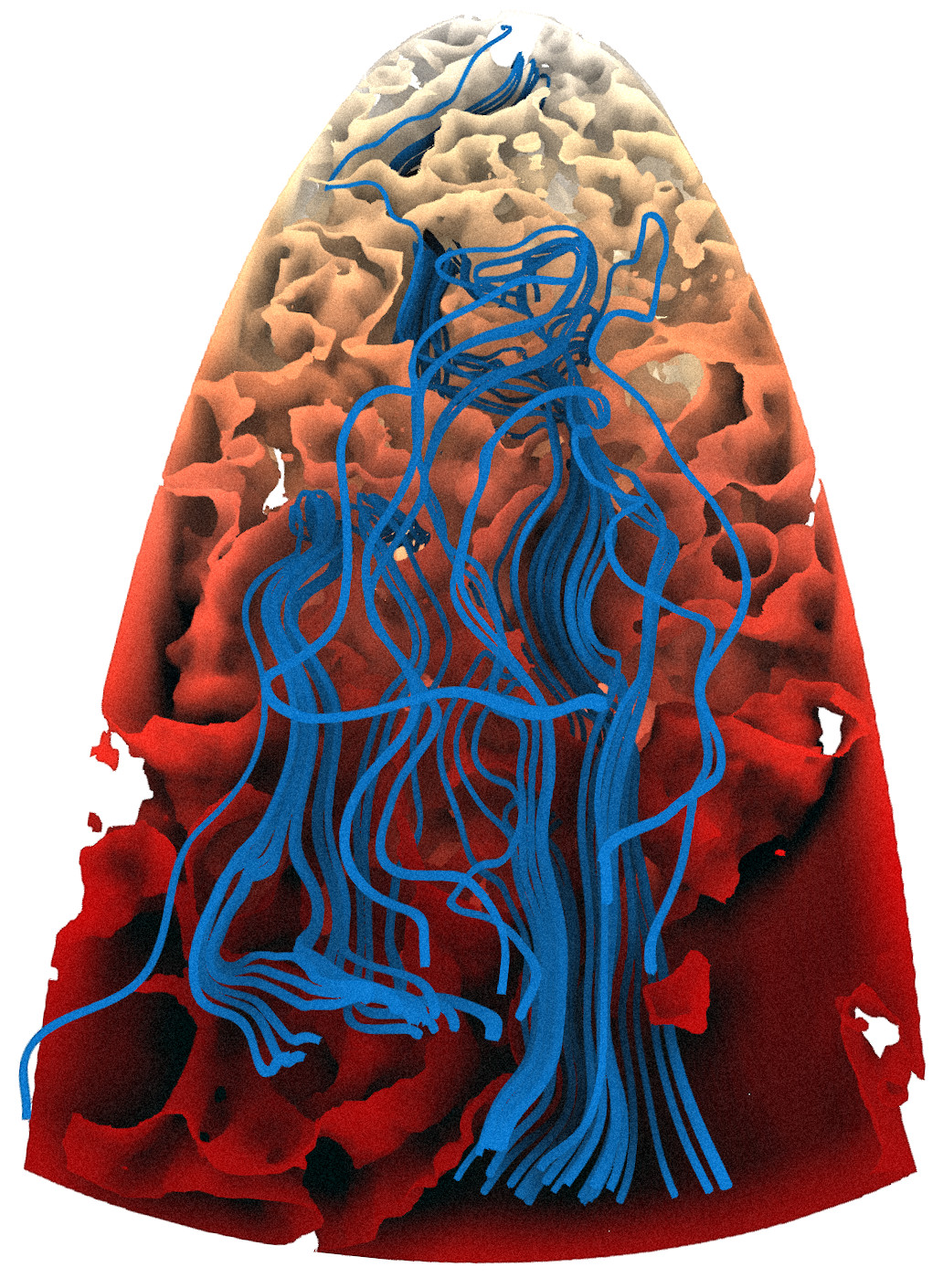}{};
            \subfigure{1}{0}{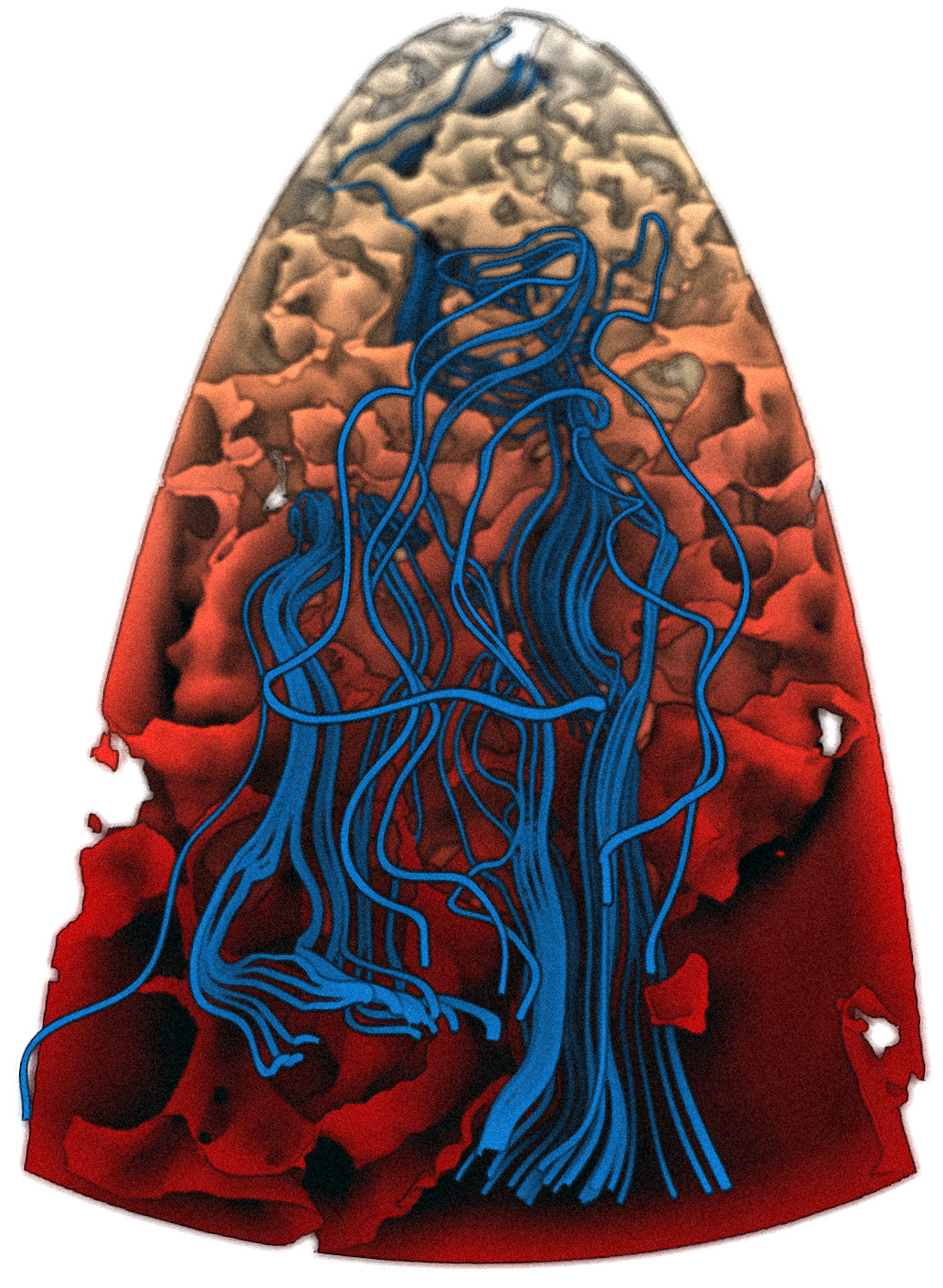}{};
        \end{tikzpicture}%
        \hspace*{-30em}\vspace*{-2em}
    \end{center}

    \caption{
        This rendering shows the path of water (blue streamlines) through a karst limestone ground sample (colored by height) taken in south Florida, USA~\cite{ttkData}.
        The corresponding Cinema database only consists of two depth images that separately depict the stone sample and the streamlines.
        Both depth buffers are composited post hoc, where the height field on the stone is reconstructed in image space.
        SSAO is used to approximate global illumination (left), and spatial perception is further improved via IBS~\cite{dntg2020} and SSDoF~\cite{mcintosh2012efficiently}~(right).\vspace*{-1em}
    }
    \label{fig_stone}
\end{figure}

\section*{Acknowledgments}
\vspace*{-0.5em}
{

\small
This work was supported by the U.S. Department of Homeland
Security under Grant Award 2017-ST-061-QA0001 and 17STQAC00001-03-03, and the
National Science Foundation Program under Award No. 1350573.
The views and conclusions contained in this document are those of the authors
and should not be interpreted as necessarily representing the official policies,
either expressed or implied, of the U.S. Department of Homeland Security.
This work was also partially supported by the German research foundation
(DFG) through~the~IRTG~2057, and by a grant from the Swedish Foundation for Strategic Research (SSF, BD15-0082), the SeRC (Swedish e-Science Research Center) and the ELLIIT environment for strategic research in Sweden.}

\bibliographystyle{abbrv-doi}

\bibliography{main}

\begin{thebibliography}{10}

\bibitem{blast}
{BLAST: High-Order Finite Element Hydrodynamics, 2020}.
\newblock \\\url{https://computing.llnl.gov/projects/blast}.

\bibitem{DevilRay}
{Devil Ray: High-Order Ray Tracer, 2020}.
\newblock \\\url{ https://github.com/LLNL/devil_ray}.

\bibitem{ahrens2015image}
J.~Ahrens, S.~Jourdain, P.~O'Leary, J.~Patchett, D.~Rogers, and M.~Petersen.
\newblock {An Image-Based Approach to Extreme Scale In Situ Visualization and
  Analysis}.
\newblock In {\em Proceedings of the International Conference for High
  Performance Computing, Networking, Storage and Analysis}, pp. 424--434, 2014.

\bibitem{aldrich2019query}
G.~Aldrich, J.~Lukasczyk, J.~D. Hyman, G.~Srinivasan, H.~Viswanathan, C.~Garth,
  H.~Leitte, J.~P. Ahrens, and B.~Hamann.
\newblock {A Query-Based Framework for Searching, Sorting, and Exploring Data
  Ensembles}.
\newblock {\em Computing in Science \& Engineering}, 22(2):64--76, 2019.

\bibitem{ayachit2015paraview}
U.~Ayachit, A.~Bauer, B.~Geveci, P.~O'Leary, K.~Moreland, N.~Fabian, and
  J.~Mauldin.
\newblock {Paraview Catalyst: Enabling In Situ Data Analysis and
  Visualization}.
\newblock In {\em Proceedings of the First Workshop on In Situ Infrastructures
  for Enabling Extreme-Scale Analysis and Visualization}, pp. 25--29, 2015.

\bibitem{bello2020matrix}
P.~D. Bello-Maldonado, T.~V. Kolev, R.~N. Rieben, and V.~Z. Tomov.
\newblock {A Matrix-Free Hyperviscosity Formulation for High-Order ALE
  Hydrodynamics}.
\newblock {\em Computers \& Fluids}, p. 104577, 2020.

\bibitem{berger2018generative}
M.~Berger, J.~Li, and J.~A. Levine.
\newblock {A generative Model for Volume Rendering}.
\newblock {\em IEEE transactions on visualization and computer graphics},
  25(4):1636--1650, 2018.

\bibitem{biedert2015}
T.~Biedert and C.~Garth.
\newblock {Contour Tree Depth Images for Large Data Visualization}.
\newblock In {\em Proceedings of the 15th Eurographics Symposium on Parallel
  Graphics and Visualization}, pp. 77--86, 2015.

\bibitem{blender2018}
{Blender Online Community}.
\newblock {Blender - A 3D Modelling and Rendering Package}.
\newblock 2018.
\newblock \url{http://www.blender.org}.

\bibitem{liang2000deferred}
{Bor-Sung Liang}, {Wen-Chang Yeh}, {Yuan-Chung Lee}, and {Chein-Wei Jen}.
\newblock {Deferred Lighting: A Computation-Efficient Approach for Real-Time
  3-D Graphics}.
\newblock In {\em 2000 IEEE International Symposium on Circuits and Systems
  (ISCAS)}, vol.~4, pp. 657--660 vol.4, 2000.

\bibitem{cohen2002three}
R.~H. Cohen, W.~P. Dannevik, A.~M. Dimits, D.~E. Eliason, A.~A. Mirin, Y.~Zhou,
  D.~H. Porter, and P.~R. Woodward.
\newblock {Three-Dimensional Simulation of a Richtmyer–Meshkov Instability
  with a Two-Scale Initial Perturbation}.
\newblock {\em Physics of Fluids}, 14(10):3692--3709, 2002.

\bibitem{deering1988triangle}
M.~Deering, S.~Winner, B.~Schediwy, C.~Duffy, and N.~Hunt.
\newblock {The Triangle Processor and Normal Vector Shader: A VLSI System for
  High Performance Graphics}.
\newblock {\em ACM Siggraph computer graphics}, 22(4):21--30, 1988.

\bibitem{dirksen2013learning}
J.~Dirksen.
\newblock {\em {Learning Three.js: The JavaScript 3D Library for WebGL}}.
\newblock Packt Publishing Ltd, 2013.

\bibitem{eichelbaum2013pointao}
S.~Eichelbaum, G.~Scheuermann, and M.~Hlawitschka.
\newblock {PointAO-Improved Ambient Occlusion for Point-based Visualization}.
\newblock In {\em EuroVis (Short Papers)}, 2013.

\bibitem{engelke2019autonomous}
W.~Engelke, K.~Lawonn, B.~Preim, and I.~Hotz.
\newblock {Autonomous Particles for Interactive Flow Visualization}.
\newblock In {\em Computer Graphics Forum}, vol.~38, pp. 248--259. Wiley Online
  Library, 2019.

\bibitem{unrealengine}
{Epic Games}.
\newblock {Unreal Engine}, 2020.
\newblock {\fontsize{7.5}{7.5}\url{https://www.unrealengine.com}}.

\bibitem{fernandes2014space}
O.~Fernandes, S.~Frey, F.~Sadlo, and T.~Ertl.
\newblock Space-time volumetric depth images for in-situ visualization.
\newblock In {\em Proceedings of IEEE 4th Symposium on Large Data Analysis and
  Visualization (LDAV)}, pp. 59--65, 2014.

\bibitem{garthJetData}
C.~Garth.
\newblock {Simulation of a Jet Flow}, 2020. doi: {{%
10\hspace{.1pt}\discretionary{.}{%
}{.}\hspace{.4pt}21227\discretionary{/}{%
}{/}qjxp\discretionary{%
}{-}{-}kc31}}


\bibitem{grottel2014megamol}
S.~Grottel, M.~Krone, C.~M{\"u}ller, G.~Reina, and T.~Ertl.
\newblock {MegaMol—A Prototyping Framework for Particle-Based Visualization}.
\newblock {\em IEEE transactions on visualization and computer graphics},
  21(2):201--214, 2014.

\bibitem{harper2012mastering}
J.~Harper.
\newblock {\em {Mastering Autodesk 3ds Max}}.
\newblock John Wiley \& Sons, 2012.

\bibitem{Huang2012}
J.~{Huang}, W.~{Pei}, C.~{Wen}, G.~{Chen}, W.~{Chen}, and H.~{Bao}.
\newblock {Output-Coherent Image-Space LIC for Surface Flow Visualization}.
\newblock In {\em 2012 IEEE Pacific Visualization Symposium}, pp. 137--144,
  2012.

\bibitem{jonsson2019inviwo}
D.~J{\"o}nsson, P.~Steneteg, E.~Sund{\'e}n, R.~Englund, S.~Kottravel, M.~Falk,
  A.~Ynnerman, I.~Hotz, and T.~Ropinski.
\newblock {Inviwo-A Visualization System with Usage Abstraction Levels}.
\newblock {\em IEEE transactions on visualization and computer graphics}, 2019.

\bibitem{openSciVisDataSets}
P.~Klacansky.
\newblock {Open Scientific Visualization Datasets, 2020}.
\newblock \\\url{https://klacansky.com/open-scivis-datasets/}.

\bibitem{larsen2017alpine}
M.~Larsen, J.~Ahrens, U.~Ayachit, E.~Brugger, H.~Childs, B.~Geveci, and
  C.~Harrison.
\newblock The alpine in situ infrastructure: Ascending from the ashes of
  strawman.
\newblock In {\em Proceedings of the In Situ Infrastructures on Enabling
  Extreme-Scale Analysis and Visualization}, pp. 42--46. 2017.

\bibitem{lottes2009fast}
T.~Lottes.
\newblock {Fast Approximate Anti-Aliasing (FXAA)}.
\newblock {\em NVIDIA Corporation, Santa Clara, CA, USA, Feb}, 2009.

\bibitem{luft2006image}
T.~Luft, C.~Colditz, and O.~Deussen.
\newblock {\em {Image Enhancement by Unsharp Masking the Depth Buffer}},
  vol.~25.
\newblock ACM, 2006.

\bibitem{dntg2020}
J.~{Lukasczyk}, C.~{Garth}, G.~H. {Weber}, T.~{Biedert}, R.~{Maciejewski}, and
  H.~{Leitte}.
\newblock {Dynamic Nested Tracking Graphs}.
\newblock {\em IEEE Transactions on Visualization and Computer Graphics},
  26(1):249--258, 2020.

\bibitem{lukasczyk2018voidga}
J.~Lukasczyk, E.~Kinner, J.~Ahrens, H.~Leitte, and C.~Garth.
\newblock {VOIDGA: A View-Approximation Oriented Image Database Generation
  Approach}.
\newblock In {\em IEEE 8th Symposium on Large Data Analysis and Visualization
  (LDAV)}, 2018.

\bibitem{marrin2011webgl}
C.~Marrin.
\newblock {Webgl Specification}.
\newblock {\em Khronos WebGL Working Group}, 2011.
\newblock
  {\fontsize{7.5}{7.5}\selectfont\url{https://www.khronos.org/registry/webgl/specs/latest/1.0/}}.

\bibitem{mcguire2012scalable}
M.~McGuire, M.~Mara, and D.~P. Luebke.
\newblock Scalable ambient obscurance.
\newblock In {\em High Performance Graphics}, pp. 97--103. Citeseer, 2012.

\bibitem{mcintosh2012efficiently}
L.~McIntosh, B.~E. Riecke, and S.~DiPaola.
\newblock Efficiently simulating the bokeh of polygonal apertures in a
  post-process depth of field shader.
\newblock In {\em Computer Graphics Forum}, vol.~31, pp. 1810--1822. Wiley
  Online Library, 2012.

\bibitem{moreland2016vtk}
K.~Moreland, C.~Sewell, W.~Usher, L.-t. Lo, J.~Meredith, D.~Pugmire, J.~Kress,
  H.~Schroots, K.-L. Ma, H.~Childs, et~al.
\newblock {VTK-m: Accelerating the Visualization Toolkit for Massively Threaded
  Architectures}.
\newblock {\em IEEE computer graphics and applications}, 36(3):48--58, 2016.

\bibitem{o2016cinema}
P.~O’Leary, J.~Ahrens, S.~Jourdain, S.~Wittenburg, D.~H. Rogers, and
  M.~Petersen.
\newblock {Cinema Image-Based In Situ Analysis and Visualization of MPAS-Ocean
  Simulations}.
\newblock {\em Parallel Computing}, 55:43--48, 2016.

\bibitem{popinet2003gerris}
S.~Popinet.
\newblock {Gerris: A Tree-Based Adaptive Solver for the Incompressible Euler
  Equations in Complex Geometries}.
\newblock {\em Journal of Computational Physics}, 190(2):572--600, 2003.
\newblock \url{http://gfs.sf.net/}.

\bibitem{ritschel2009approximating}
T.~Ritschel, T.~Grosch, and H.-P. Seidel.
\newblock {Approximating Dynamic Global Illumination in Image Space}.
\newblock In {\em Proceedings of the 2009 Symposium on Interactive 3D Graphics
  and Games}, pp. 75--82, 2009.

\bibitem{rogers2018cinemaspec}
D.~Rogers, J.~Woodring, J.~Ahrens, J.~Patchett, and J.~Lukasczyk.
\newblock {Cinema Database Specification - Dietrich Release v1.2}.
\newblock Technical Report LA-UR-17-25072, Los Alamos National Laboratory,
  2018.

\bibitem{vtkbook}
W.~J. Schroeder, K.~Martin, and W.~E. Lorensen.
\newblock {\em {The Visualization Toolkit: An Object Oriented Approach to 3D
  Graphics}}.
\newblock {Kitware, Inc.}, 2004.

\bibitem{ttk2017}
J.~Tierny, G.~Favelier, J.~A. Levine, C.~Gueunet, and M.~Michaux.
\newblock {The Topology ToolKit}.
\newblock {\em {IEEE Transactions on Visualization and Computer Graphics}},
  2017.
\newblock \url{https://topology-tool-kit.github.io/}.

\bibitem{tikhonova2010explorable}
A.~{Tikhonova}, C.~D. {Correa}, and K.~{Ma}.
\newblock {Explorable Images for Visualizing Volume Data}.
\newblock In {\em 2010 IEEE Pacific Visualization Symposium (PacificVis)}, pp.
  177--184, 2010.

\bibitem{ttkData}
{TTK Contributers}.
\newblock {\textit{TTK Data Repository}, 2020}.
\newblock \\\url{https://github.com/topology-tool-kit/ttk-data}.

\bibitem{unityengine}
{Unity Technologies}.
\newblock {Unity Game Engine}.
\newblock 2020.
\newblock {\fontsize{7.5}{7.5}\selectfont\url{https://unity.com}}.

\bibitem{embree2014}
I.~Wald, S.~Woop, C.~Benthin, G.~S. Johnson, and M.~Ernst.
\newblock {Embree: A Kernel Framework for Efficient CPU Ray Tracing}.
\newblock {\em ACM Trans. Graph.}, 33(4), July 2014.

\bibitem{weinkauf10c}
T.~Weinkauf and H.~Theisel.
\newblock {Streak Lines as Tangent Curves of a Derived Vector Field}.
\newblock {\em IEEE Transactions on Visualization and Computer Graphics
  (Proceedings Visualization 2010)}, 16(6):1225--1234, 2010.

\end{thebibliography}

\end{document}